# High-Temperature Superconductivity in Th-H System at Pressure Conditions

Alexander G. Kvashnin, [1,2,*] Dmitrii V. Semenok, [1,2] Ivan A. Kruglov, [2,3] Izabela A. Wrona, [4] Artem R. Oganov, [1,2,3,5*]

[1] Skolkovo Institute of Science and Technology, Skolkovo Innovation Center 143026, 3 Nobel Street, Moscow, Russian Federation
[2] Moscow Institute of Physics and Technology, 141700, 9 Institutsky lane, Dolgoprudny, Russian Federation
[3] Dukhov Research Institute of Automatics (VNIIA), Moscow 127055, Russian Federation
[4] Institute of Physics, Jan Dlugosz University in Czestochowa, Ave. Armii Krajowej 13/15, 42-200 Czestochowa, Poland
[5] International Center for Materials Discovery, Northwestern Polytechnical University, Xi'an, 710072, China

Corresponding Author
*A.G. Kvashnin, E-mail: A.Kvashnin@skoltech.ru
*A.R. Oganov, E-mail: A.Oganov@skoltech.ru

**ABSTRACT.** New stable thorium decahydride $Fm\bar{3}m$-ThH$_{10}$, a record high-temperature superconductor with $T_C$ up to 241 K (-32ºC), critical field $H_C$ up to 71 T and superconducting gap $\Delta_0 = 52$ meV at 80-100 GPa was predicted by evolutionary algorithm USPEX. Another phase $P2_1/c$-ThH$_7$ was found to be a high-temperature superconductor with $T_C \sim 65$ K. Analysis of superconducting state was performed within Eliashberg formalism and the dependencies of $H_C(T)$, $\Delta(T)$, $T_C(P)$ together with jump in the specific heat at critical temperature were calculated. Several other new thorium hydrides were predicted to be stable under pressure including ThH$_3$, Th$_3$H$_{10}$, ThH$_4$, ThH$_6$. Thorium (which has $s^2d^2$ electronic configuration) forms high-$T_C$ polyhydrides similar to those formed by $s^2d^1$ metals (Y-La-Ac). Thorium – is the next member in Mg-Ca-Sc-Y-La-Ac family of elements forming high-$T_C$ superconducting hydrides.

## Introduction

Recent outstanding experimental observation of previously predicted [1] superconductivity in the LaH$_{10+x}$ by Drozdov et al. [2] and Somayazulu et al. [3] with record temperatures 215-245 K (150-192 GPa) inspired us to search for new high-$T_C$ superconductors among Th-H system where one uniquely stable compound was found.

Thorium is a weakly radioactive element, its most stable isotope [232]Th having a half-life of 14 billion years. Thorium belongs to the actinide series, but its physical and chemical properties are mainly dictated by $6d$ electrons, similar to transition metals Ti, Hf, or Zr [4]. Bulk thorium has fcc structure ($Fm\bar{3}m$ space group) in contrast to other elements of actinide series [5] and is a weak conventional (phonon-mediated) type-I superconductor with $T_C = 1.374$ K [5,6] and critical magnetic field $H_C = 15.92$ mT [7,5]. Theoretical predictions of the electron–phonon coupling have been reported to be in good agreement with experimental data [8–10]. Chemistry of actinides and their compounds is a relatively unpopular topic, but many potentially interesting materials can be discovered there. Here we show that unique high-temperature superconductors are predicted among thorium hydrides.

Thorium reacts with hydrogen to form hydrides ThH$_2$ and Th$_4$H$_{15}$, the latter being the first known superconducting hydride with metallic conductivity at normal conditions, with superconducting $T_C = 7.5$–8 K [11,12]. They are chemically active and sensitive to air and moisture [13]. Stability and electronic properties of ThH$_2$ and Th$_4$H$_{15}$ were studied theoretically [14,15]. Investigations of superconducting properties in hydrides and deuterides of thorium with H- or D-to-metal atom ratios of 3.6 to 3.65 revealed the absence of the isotope effect [16], either due to predominant anharmonic contribution or non-phonon pairing mechanisms.

Recent experimental and theoretical studies show that some of hydrogen-rich hydrides are potential high-temperature superconductors under high pressure [1,17–25]. Hydrogen sulfide stands out among the rest. Using evolutionary algorithm USPEX [26–28] $H_3S$ in the trigonal *R3m* and cubic *Im$\bar{3}$m* phases was predicted at high pressures (above 120 GPa), and was predicted to be a high-temperature superconductor with $T_C$ in the range 191-204 K at 200 GPa for the *Im$\bar{3}$m* phase [19]. Experimental work [29] confirmed the existence of superconducting $H_3S$ with $T_C$ = 203 K at 155 GPa, and its crystal structure was confirmed to be *Im$\bar{3}$m* [22,30]. Recent predictions of potential high-temperature superconductivity in lanthanum, yttrium, actinium and uranium hydrides ($LaH_{10}$ and $YH_{10}$ [1], $AcH_{10}$ [31] and $UH_7$, $UH_9$ [25]) and recent experimental syntheses of $LaH_{10+x}$ [32], $CeH_9$ [33], $CrH_2$ and $Cr_2H_3$ [34] show that transition metal hydrides are extremely promising. Thus, we decided to perform a systematic evolutionary search for new phases in the Th-H system under pressure. As we show below, one of the new hydrides is predicted to be a unique high-temperature superconductor.

## Computational Methodology

Evolutionary algorithm USPEX [26–28] is a powerful tool for predicting thermodynamically stable compounds of given elements at a given pressure. We performed variable-composition searches in the Th-H system at pressures 0, 50, 100, 150 and 200 GPa. The first generation (120 structures) was created using random symmetric generator, while all subsequent generations contained 20% random structures, and 80% created using heredity, softmutation and transmutation operators. Here, evolutionary searches were combined with structure relaxations using density functional theory (DFT) [35,36] within the generalized gradient approximation (Perdew-Burke-Ernzerhof functional) [37], and the projector augmented wave method [38,39] as implemented in the VASP code [40–42]. Plane wave kinetic energy cutoff was set to 600 eV and the Brillouin zone was sampled using Γ-centered k-points meshes with resolution $2\pi \times 0.05$ Å$^{-1}$.

In order to establish stability fields of the predicted phases, we recalculated their enthalpies with increased precision at various pressures with a smaller pressure increment (5-10 GPa), recalculating the thermodynamic convex hull (Maxwell construction) at each pressure. By definition, a thermodynamically stable phase has lower Gibbs free energy (or, at zero Kelvin, lower enthalpy) than any phase or phase assemblage of the same composition. Thus, phases that are located on the convex hull are the ones stable at given pressure. Stable structures of elemental Th and H were taken from USPEX calculations and from Refs. [43] and [44].

Calculations of superconducting $T_C$ were carried out using QUANTUM ESPRESSO (QE) package [45]. Phonon frequencies and electron-phonon coupling (EPC) coefficients were computed using density-functional perturbation theory [46], employing plane-wave pseudopotential method and Perdew-Burke-Ernzerhof exchange-correlation functional [37]. Convergence tests showed that 120 Ry is a suitable kinetic energy cutoff for the plane wave basis set. Electronic band structures of newly found thorium hydrides were calculated using both VASP and QE, and demonstrated good consistency. Comparison of phonon densities of states calculated using finite displacement method (VASP and PHONOPY [47,48]) and density-functional perturbation theory (QE) showed perfect agreement between these methods.

Critical temperature $T_C$ was calculated from the Eliashberg equation [49], which is based on the Fröhlich Hamiltonian $\hat{H} = \hat{H}_e + \hat{H}_{ph} + \sum_{k,q,j} g_{k+q,k}^{q,j} \hat{c}_{k+q}^+ \hat{c}_k (\hat{b}_{-q,j}^+ + \hat{b}_{q,j})$, where $c^+$, $b^+$ are creation operators of electrons and phonons, respectively. Matrix elements of electron-phonon interaction $g_{k+q,k}^{q,j}$ calculated in the harmonic approximation in Quantum ESPRESSO package can be defined as $g_{k+q,k}^{q,j} = \sqrt{\dfrac{\hbar}{2M\omega_{q,j}}} \int \psi_k^*(r) \cdot \left\{ \dfrac{dV_{scf}}{d\vec{u}_q} \cdot \dfrac{\vec{u}_q}{|\vec{u}_q|} \right\} \cdot \psi_{k+q}(r) d^3r$, where $u_q$ is displacement of an atom with mass $M$ in the phonon mode $q,j$. Within the framework of Gor'kov and Migdal approach [50,51] the correction to the electron Green's function $\Sigma(\vec{k},\omega) = G_0^{-1}(\vec{k},\omega) - G^{-1}(\vec{k},\omega)$ caused by interaction can be calculated by taking into account only the first terms of expansion of electron-phonon interaction in $\omega_{ph}/E_F$. As a result, it will lead to integral Eliashberg equations [49]. These equations can be solved by iterative self-consistent method for the real part of the order parameter $\Delta(T, \omega)$ (superconducting gap) and the renormalization wave function $Z(T, \omega)$ [52] (see Supporting Information).

In our *ab initio* calculations of the electron-phonon coupling (EPC) parameter $\lambda$, the first Brillouin zone was sampled using 6×6×6 *q*-points mesh, and a denser 24×24×24 *k*-points mesh (with Gaussian smearing and σ = 0.025 Ry, which approximates the zero-width limits in the calculation of λ). In addition to the full solution of the Eliashberg equations, the simplified formulas "full" – Allen-Dynes and "short" – modified McMillan equation [53]. The "full" Allen-Dynes equation for calculating $T_C$ has the following form [53]:

$$T_C = \omega_{log} \frac{f_1 f_2}{1.2} \exp\left( \frac{-1.04(1+\lambda)}{\lambda - \mu^* - 0.62\lambda\mu^*} \right) \qquad (1)$$

with

$$f_1 f_2 = \sqrt[3]{1 + \left(\frac{\lambda}{2.46(1+3.8\mu^*)}\right)^{\frac{3}{2}}} \cdot \left(1 - \frac{\lambda^2(1 - \omega_2/\omega_{log})}{\lambda^2 + 3.312(1+6.3\mu^*)^2}\right), \qquad (2)$$

while the modified McMillan equation has the form but with $f_1 f_2 = 1$.

The EPC constant $\lambda$, logarithmic average frequency $\omega_{log}$ and mean square frequency $\omega_2$ were calculated as:

$$\lambda = \int_0^{\omega_{max}} \frac{2 \cdot \alpha^2 F(\omega)}{\omega} d\omega \qquad (3)$$

and

$$\omega_{log} = \exp\left(\frac{2}{\lambda} \int_0^{\omega_{max}} \frac{d\omega}{\omega} \alpha^2 F(\omega) \ln(\omega)\right), \omega_2 = \sqrt{\frac{1}{\lambda} \int_0^{\omega_{max}} \left[\frac{2\alpha^2 F(\omega)}{\omega}\right] \omega^2 d\omega} \qquad (4)$$

and $\mu^*$ is the Coulomb pseudopotential, for which we used widely accepted lower and upper bound values of 0.10 and 0.15.

Sommerfeld constant was found as

$$\gamma = \frac{2}{3}\pi^2 k_B^2 N(0)(1+\lambda), \qquad (5)$$

and used to estimate the upper critical magnetic field and the superconductive gap in $Fm\bar{3}m$-ThH$_{10}$ at 200 and 300 GPa by well-known semi-empirical equations of the BCS theory (see Ref. [54], equations 4.1 and 5.11), working satisfactorily for $T_C/\omega_{\log} < 0.25$ (see Table S3):

$$\frac{\gamma T_C^2}{H_C^2(0)} = 0.168 \left[ 1 - 12.2 \left( \frac{T_C}{\omega_{\log}} \right)^2 \ln \left( \frac{\omega_{\log}}{3T_C} \right) \right] \quad (6)$$

$$\frac{2\Delta(0)}{k_B T_C} = 3.53 \left[ 1 + 12.5 \left( \frac{T_C}{\omega_{\log}} \right)^2 \ln \left( \frac{\omega_{\log}}{2T_C} \right) \right] \quad (7)$$

## Results

We built the composition-pressure phase diagram (see Fig. 1), which shows pressure ranges of stability of all found phases at different pressures. As shown in Fig. 1, our calculations correctly reproduce stability of ThH$_2$ and Th$_4$H$_{15}$ and predict 8 new stable phases namely $R\bar{3}m$-ThH$_3$, $Immm$-Th$_3$H$_{10}$, $Pnma$-ThH$_4$, $P321$-ThH$_4$, $I4/mmm$-ThH$_4$, $Cmc2_1$-ThH$_6$, $P2_1/c$-ThH$_7$ and $Fm\bar{3}m$-ThH$_{10}$.

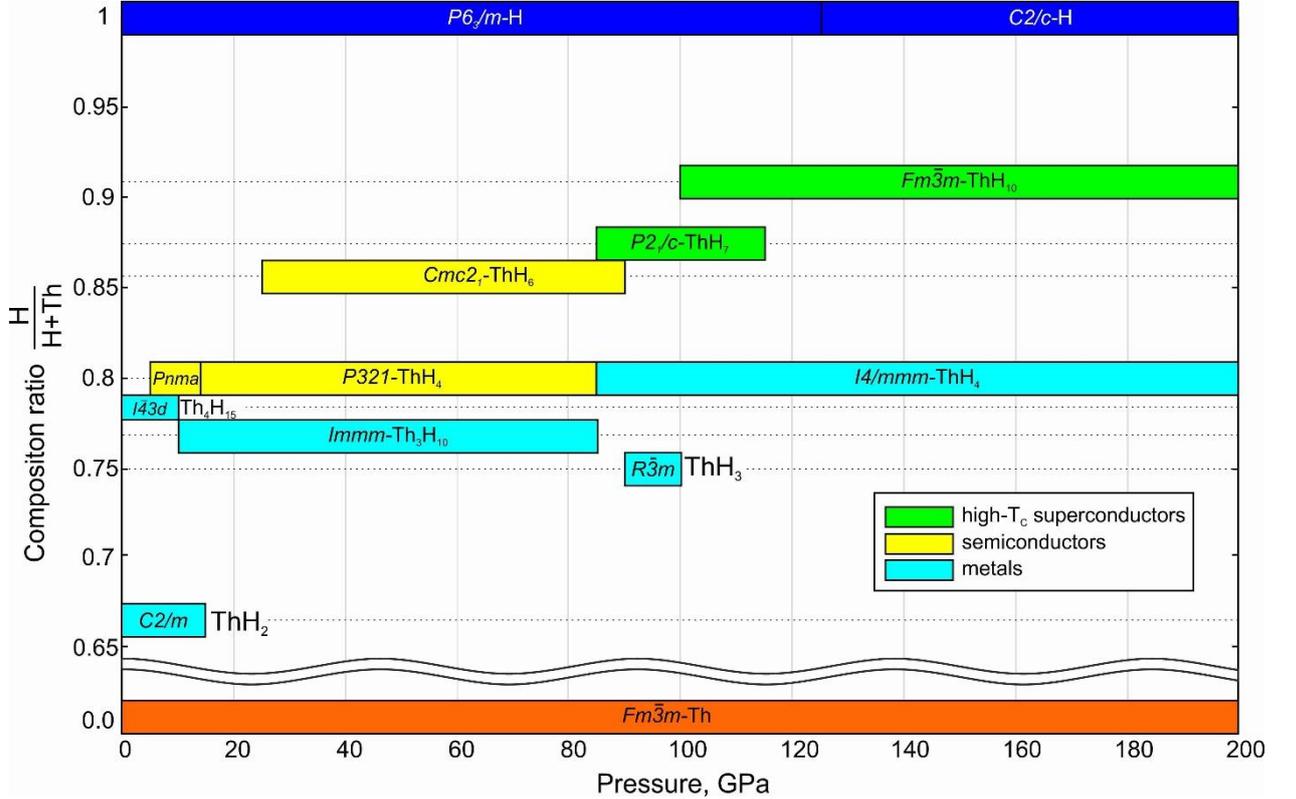

Fig. 1 Pressure-composition phase diagram of the Th-H system. Yellow, blue and green indicate semiconducting, normal metallic and high-$T_C$ superconducting phases, respectively.

Detailed information on crystal structures of the predicted phases is summarized in Table S1 (see Supporting Information). It is clearly seen from our calculations that at the pressure of 5 GPa the new semiconducting $Pnma$-ThH$_4$ phase becomes stable, while increasing pressure leads to the phase transition to another semiconducting $P321$-ThH$_4$ at about 15 GPa. Further increase of pressure leads to the $P321 \rightarrow I4/mmm$ phase transition at 90 GPa. At the pressure of 10 GPa, the $Immm$-Th$_3$H$_{10}$ phase becomes thermodynamically stable and remains stable up to 90 GPa. $Cmc2_1$-ThH$_6$ phase appears on the phase diagram in the pressure range 25-90 GPa. The $R\bar{3}m$-ThH$_3$ phase is stable in a very narrow pressure range from 90 to 100 GPa. At pressures higher than 100 GPa only 3 phases are stable, namely $I4/mmm$-ThH$_4$, $P2_1/c$-ThH$_7$ and $Fm\bar{3}m$-ThH$_{10}$. Importantly, among the known high-$T_C$ superconducting hydrides like LaH$_{10}$ [1], YH$_{10}$ [1], AcH$_{10}$ [31], thorium decahydride is unique due to the lowest stabilization pressure of 100 GPa, compared to 170 GPa for LaH$_{10+x}$ [32], 250 GPa for YH$_{10}$ [1] and 200 GPa for AcH$_{10}$ [31]. Scandium decahydride is unstable and transforms to ScH$_9$ [55]. The $I4/mmm$-ThH$_4$ and $Fm\bar{3}m$-ThH$_{10}$ phases are stable at least up to 200 GPa, while $P2_1/c$-ThH$_7$ is unstable at pressures above 120 GPa.

It is important to note that three of predicted Th-H phases are semiconductors (yellow color in Fig. 1) with the DFT band gaps around 0.6-0.9 eV; systematic underestimation of band gaps by DFT calculations is well known. The rest of predicted phases $R\bar{3}m$-ThH$_3$, $Immm$-Th$_3$H$_{10}$, $I4/mmm$-ThH$_4$, $P2_1/c$-ThH$_7$ and $Fm\bar{3}m$-ThH$_{10}$ are metallic and superconducting. Crystal structure of ThH$_{10}$ is shown in Fig. 2. The atoms occupy large 24-coordinatd voids in the sodalite-type framework of hydrogens. The $P2_1/c$-ThH$_7$ phase has thorium atoms forming a bcc sublattice; each thorium atom has irregular coordination by hydrogen atoms. (see Fig. 2). Thorium coordination number is remarkably high (23), distances are in the range from 2.03 to 2.23 Å at 100 GPa. We observe high DOS at the Fermi level, which also should be favorable for superconducting properties (see Fig. 2).

Phonon calculations confirmed that none of the newly predicted superconducting phases have imaginary phonon frequencies in their predicted ranges of thermodynamic stability (see Fig. 2 and Fig. S1). Calculated partial phonon densities of states of $Fm\bar{3}m$-ThH$_{10}$ show the presence of three major contributions. Two peaks at ~5 THz correspond to vibrations of thorium atoms inside the hydrogen cage. The next sets of peaks (shown in Fig. 2 by green color) correspond to vibrations of hydrogen atoms located between the hydrogen cubes (H1 type with H-H distance of 1.297 Å). High-frequency region corresponds to hydrogen atoms in the cubes (H2 type, blue color) with H-H distance of 1.149 Å. Structures of $R\bar{3}m$-ThH$_3$, $Immm$-Th$_3$H$_{10}$, $I4/mmm$-ThH$_4$ and $P2_1/c$-ThH$_7$ phases are shown in Fig. S1 (see Supporting Information).

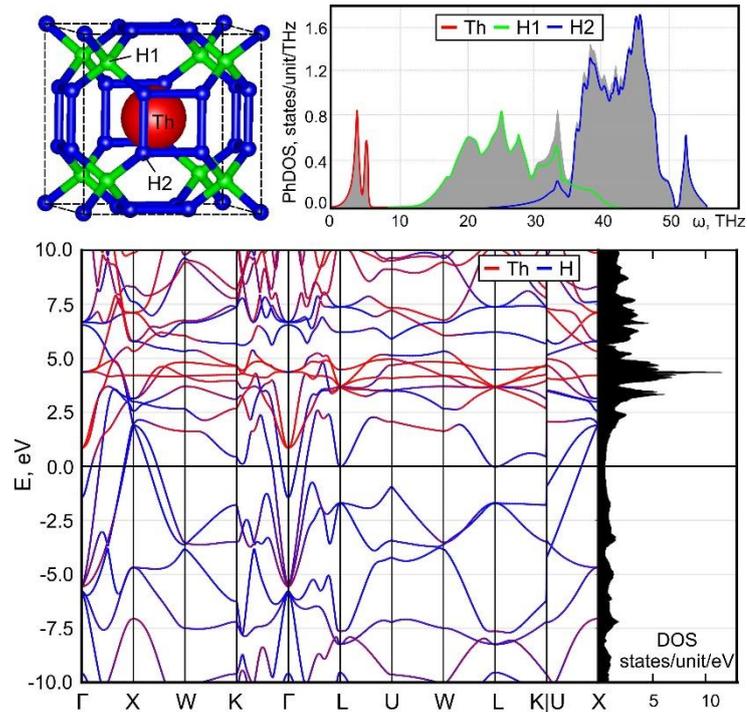

Fig. 2 Crystal structure, partial phonon density of states, electronic band structure and density of states for predicted $Fm\bar{3}m$-ThH$_{10}$ at 100 GPa. Colors in phonon DOS correspond to colors of atoms in the drawing of the crystal. Crystal structures of predicted phases were visualized using VESTA software [56].

Calculated superconducting properties of thorium hydrides are given in Table 1. The following new phases are metallic and potentially superconducting: $R\bar{3}m$-ThH$_3$, $Immm$-Th$_3$H$_{10}$, $I4/mmm$-ThH$_4$, $P2_1/c$-ThH$_7$ and $Fm\bar{3}m$-ThH$_{10}$. For $Fm\bar{3}m$-ThH$_{10}$ at 100 GPa, the EPC coefficient $\lambda = 2.50$ and eq. (1) with adjustment factors f$_1$ and f$_2$ were used for calculating $T_C$. Our calculations indicate that $\alpha^2F(\omega)$ Eliashberg function of ThH$_{10}$ at 100 GPa has Nb-type behavior with two "hills". For this type of $\alpha^2F(\omega)$ function the numerical solution of the Eliashberg equations may be approximated by the full Allen-Dynes formula [53]. Transition temperature was calculated to be in a range from 177 to 221 K (depending on the exact $\mu^*$ value). Such high values of EPC coefficient and T$_C$ caused by high PhDOS intensity and localization of phonon density in a narrow frequency region around 45 THz, which corresponds to vibrations of hydrogen atoms weakly bonded to other hydrogen atoms with H-H distance of 1.149 Å (see Fig. 2).

**Table 1.** Predicted superconducting properties of thorium hydrides. $T_C$ values are given for $\mu^*$ equal to 0.1, while $T_C$ in brackets are for $\mu^* = 0.15$. $T_C$ was calculated using Allen-Dynes (*A-D*), McMillan (*McM*) equations and numerical solution of Eliashberg equation (*E*).

| Phase | P, GPa | $\lambda$ | $N_f$ states/(A$^3\cdot$Ry) | $\omega_{log}$, K | $T_C$(McM), K | $T_C$(A-D), K | $T_C$(E), K |
|---|---|---|---|---|---|---|---|
| $R\bar{3}m$-ThH$_3$ | 100 | 0.11 | 0.42 | 1664 | ~ 0 | ~ 0 | ~ 0 |
| $Immm$-Th$_3$H$_{10}$ | 10 | 0.48 | 0.48 | 379 | 3.8 (1.2) | 3.9 (1.2) | 3.9 (1.2) |
| $I4/mmm$-ThH$_4$ | 85 | 0.36 | 0.28 | 1003 | 2.97 (0.5) | 3.0 (0.5) | 3.0 (0.5) |
| $P2_1/c$-ThH$_7$ | 100 | 0.84 | 0.23 | 1192 | 61.4 (43.4) | 64.8 (45.3) | 62 (46) |
| $Fm\bar{3}m$-ThH$_{10}$ | 100 | 2.50 | 0.215 | 1073 | 176.8 (160.3) | 221.1 (193.9) | 241 (220) |
|  | 200 | 1.35 | 0.227 | 1627 | 166.3 (139.4) | 182.6 (150.5) | 228 (205) |
|  | 300 | 1.11 | 0.23 | 1775 | 144.2 (114.2) | 155.4 (121.4) | 201 (174) |

For the most interesting superconducting hydride, $Fm\bar{3}m$-ThH$_{10}$, an accurate numerical solution of Eliashberg equations was found at the lowest pressure of its stability (100 GPa) for $\mu^* = 0.1$ and 0.15 (see Supporting Information). In Fig. 3a the form of the order parameter for selected values of temperature (on imaginary axis) is shown. The superconductivity order parameter always takes the largest value for $m = 1$. The maximum value of order parameter decreases when the temperature rises and finally vanishes at the critical temperature. The influence of temperature on the maximum value of the order parameter is presented in Fig. 3b. This function can be reproduced with the help of the phenomenological formula:

$$\Delta_{m=1}(T) = \Delta_{m=1}(T_0)\sqrt{1-\left(\frac{T}{T_C}\right)^k}, \qquad (8)$$

where the value of parameter $k = 3.33$ for ThH$_{10}$ was estimated based on numerical results of $\Delta_{m=1}(T)$. In the framework of BCS theory parameter $k$ has constant value, the same for all compounds: $k_{BCS} = 3$ [57]

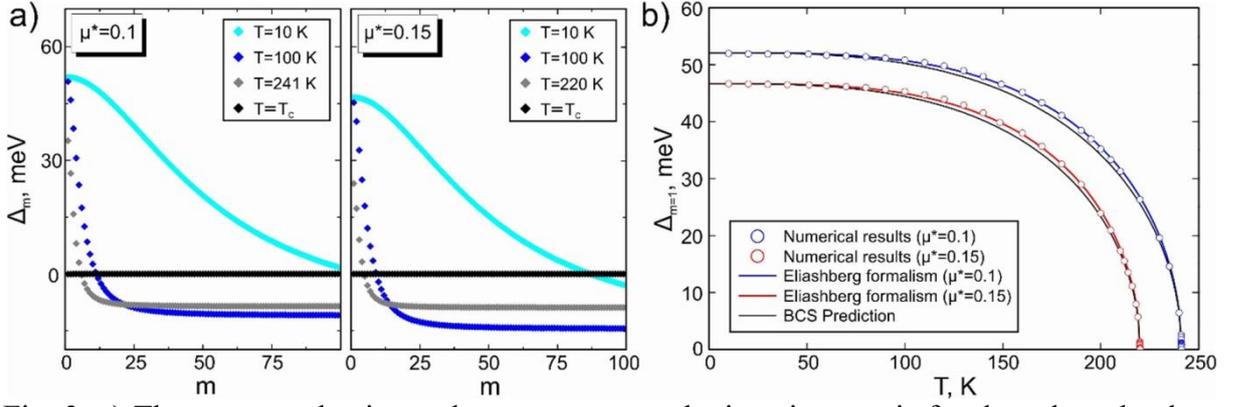

Fig. 3. a) The superconducting order parameter on the imaginary axis for the selected values of temperature calculated at $\mu^*$ equals 0.1 and 0.15; b) The temperature dependence of the maximum value of the order parameter. The lines are obtained by using equation eq. (8).

**Table 2**. Parameters of superconductive state in $Fm\bar{3}m$-ThH$_{10}$ at 100 GPa. Here $T_0 = 10$ K (initial temperature, below this temperature the solutions are not stable), $m_e$ is electron band mass, $m_e^*$ is renormalized electron mass, γ - is Sommerfeld constant.

| Parameter | Value ($\mu^* = 0.1$) | Value ($\mu^* = 0.15$) |
|---|---|---|
| $T_C$, K | 241.2 | 220.0 |
| $\Delta_{m=1}(T_0)$, meV | 52.0 | 46.7 |
| $m_e^*(T_C)$ | $3.63 m_e$ | $3.73 m_e$ |
| $H_C(T_0)$, T | 71 | 64.5 |
| $\Delta C/T_C$, mJ/mol·K$^2$ | 32.9 | 32.9 |
| γ, mJ/mol·K$^2$ | 11.1 | 11.0 |
| $R_\Delta = 2\Delta(0)/k_B T_C$ | 5.0 | 4.9 |

The high values of the wave function renormalization factor ($Z$) are related to the significant strong-coupling effects. The function of order parameter on the real axis, for selected values of temperature has been presented in Fig. S3 (see Supporting Information). It should be noted that the function takes the complex values, however only the real part is non-zero for low values of frequency, which means the infinite lifetime of Cooper pairs [58]. For higher frequencies, the damping effects become significant, which is related to the phonon emission and absorption. Complicated dependences of $Re[\Delta(\omega)]$ and $Im[\Delta(\omega)]$ for the higher frequencies (especially at $T = T_0$) are related to the shape of Eliashberg's function, which has been drawn in the background in Fig. S3. Both the real and imaginary parts are approaching zero at the critical temperature, which is related to the disappearance of superconducting properties. From the physical point of view the most important is the value corresponding to zero frequency, because it specifies the superconducting energy gap and allow one to calculate the value of ratio $R_\Delta = 2\Delta(0)/k_B T_C$ (where $\Delta(0) = Re[\Delta(\omega = 0)]_{T=T_0}$), which is a universal constant of BCS theory (3.53, compare with Table 2).

The free energy difference between the superconducting and normal state has been calculated using the following equation [54]:

$$\Delta F = -2\pi k_B T N_f \sum_{m=1}^{M} \left(\sqrt{\omega^2 + \Delta_m^2} - |\omega_m|\right) \cdot \left(Z_m^S - Z_m^N \frac{|\omega_m|}{\sqrt{\omega^2 + \Delta_m^2}}\right) \quad (9)$$

where $Z^S$ and $Z^N$ are the wave function renormalization factors for superconducting and normal states, respectively, $N_f = \rho(0)$ is the electron density of states at the Fermi level.

As a result, the maximum value of the superconducting gap was defined as 52 meV (see Table 2, Fig. 3), which exceeds that for H$_3$S (43 meV [59]). This value for ThH$_{10}$ is lower than that for LaH$_{10}$, YH$_{10}$ [1] and MgH$_6$ [60]. A thermodynamic critical field was defined as $H_c = \sqrt{8\pi(-\Delta F)}$ and equals to 64-71 T, which is close to the experimental values for sulfur hydride H$_3$S at 150 GPa ($H_C \sim$ 60-80 T) [29], but much higher than for MgB$_2$ (8 T). A jump in the specific heat at superconducting transition was calculated as $\Delta C(T) = -T \cdot d^2 \Delta F/dT^2$ and equals 7.93 J/mol·K (see Supporting Information).

For $P2_1/c$-ThH$_7$ at 100 GPa, we find a much lower EPC coefficient of 0.84, resulting in a much lower (but still high) $T_C$ in the range 43-61 K (see Table 1). At the same time, $R\bar{3}m$-ThH$_3$ cannot be considered as superconductor due to the very low EPC coefficient of 0.11 leading to $T_C$ close to 0 K. Calculated $T_C$ of the low-pressure metallic $Immm$-Th$_3$H$_{10}$ phase is 4 K and is caused by weak electron-phonon interaction ($\lambda$ = 0.48, $\omega_{log}$ = 380 K, see Table 1). The $I4/mmm$-ThH$_4$ phase shows low $T_C$ of ~ 3 K along with weak electron-phonon coupling ($\lambda$ = 0.36).

It is interesting metal atoms in predicted high-temperature superconducting hydrides of lanthanum and yttrium (LaH$_{10}$ and YH$_{10}$) [1], actinium (AcH$_{10}$) [31], scandium (ScH$_9$) [55] and thorium (ThH$_{10}$) contain almost empty d- and f-shells. Namely, Sc, Y, La, Ac all $d^1$-elements, close neighbors of $d^0$-elements, e.g. Mg (MgH$_6$) [61], Ca (CaH$_6$) [62]. Now one can see that $d^2$-element (Th) forms the same sodalite-like cubic ThH$_{10}$ and has remarkably high T$_C$ ~ 241 K. These facts allow us to formulate a rule that $d^1$, $d^2$-metals as well as $d^0$-metals from II group with minimal number of $f$-electrons will be the most promising candidates for discovering high-temperature superconductors.

As $Fm\bar{3}m$-ThH$_{10}$ phase is stable in a wide range of pressures, we determined the main parameters responsible for superconducting properties as a function of pressure (see Fig. 4). We calculated the Eliashberg spectral function $\alpha^2F(\omega)$ as a function of pressure (see Fig. 4a). One can note that increasing pressure leads to shifting of $\alpha^2F(\omega)$ function to higher frequencies. The EPC coefficient decreases with pressure, while $\omega_{log}$ increases as one can see in Fig. 4b.

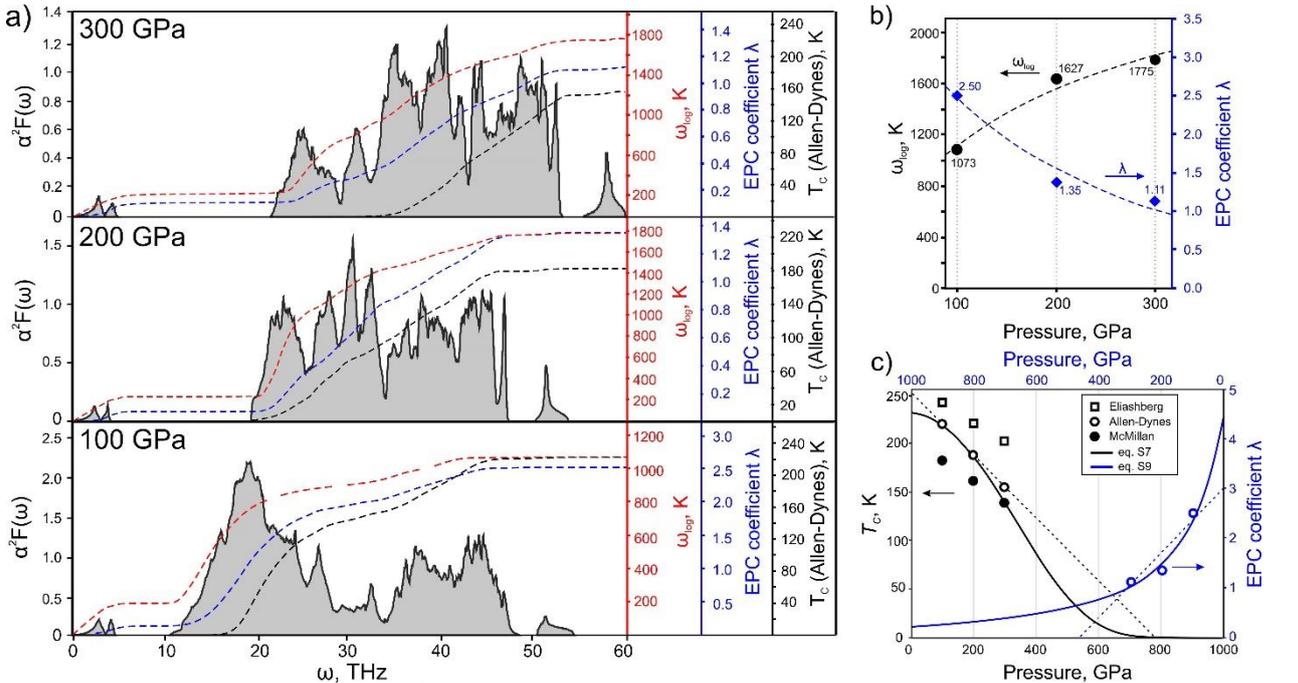

Fig. 4. a) Eliashberg function $\alpha^2F(\omega)$ as a function of pressure from 100 to 300 GPa, b) $\omega_{log}$ and EPC coefficient $\lambda$, c) critical transition temperature ($T_C$) of $Fm\bar{3}m$-ThH$_{10}$ as a function of pressure.

As one can see from Fig. 4c, the dependence of $T_C$ of ThH$_{10}$ on pressure is monotonic and non-linear, and can be described by eq. (S7). The EPC coefficient calculated using analytical eq. (S9) also decreases nonlinearly with pressure. These results can be analyzed by the well-known empirical formula for low-$T_C$ superconductors [63,64] $-ln\left(\frac{T_C}{\omega_{log}}\right) = Cv^{-\varphi}, C > 0$, where $v$ is unit cell volume (see Supporting Information for details). The computed average pressure coefficient $dT_C(E)/dP$ = -0.20 K/GPa ($\mu^*$ = 0.1) is lower than the value for CaH$_6$ (-0.33 K/GPa) [62].

Theoretically predicted LaH$_{10}$ [1] (with the same structure type as our ThH$_{10}$) was recently synthesized experimentally [53], which lends further confidence in structure prediction methods. However, experimental synthesis was achieved at a much lower pressure than predicted (170 GPa vs 250 GPa), and it is possible that our ThH$_{10}$ can also be synthesized at lower pressures than predicted by theory (i.e. below 100 GPa).

## Conclusions

In conclusion, we studied Th-H system using global optimization algorithm USPEX in order to predict new superconducting thorium hydrides, exploring pressures up to 200 GPa. Two new high-temperature superconductors were predicted, namely $Fm\bar{3}m$-ThH$_{10}$ and $P2_1/c$-ThH$_7$. $Fm\bar{3}m$-ThH$_{10}$ is predicted to be superconducting with $T_C$ in a range 220-241 K at 100 GPa, superconducting gap 47-52 meV, thermodynamic critical magnetic field 64-71 T and specific heat jump $\Delta C(T_C)/T_C$ equals 32.9 mJ/mol·K$^2$ at 100 GPa. ThH$_{10}$ is one of the highest-$T_C$ superconductors discovered so far. This study completes the investigation of a series of high-$T_C$ superconducting hydrides MgH$_6$, CaH$_6$, ScH$_9$, YH$_{10}$, LaH$_{10}$ and AcH$_{10}$. Thus, a reasonable step after exhausting of all promising binary hydrides is to pay more attention to systems of higher complexity, primarily to ternary A-B-H systems.

## Acknowledgements

The work was supported by Russian Science Foundation (№ 16-13-10459). Calculations were performed on the Rurik supercomputer at MIPT.

# Supporting Information
# High-Temperature Superconductivity in Th-H System at Pressure Conditions

Alexander G. Kvashnin, [1,2] Dmitrii V. Semenok, [1,2] Ivan A. Kruglov, [2,3] Izabela A. Wrona, [4] Artem R. Oganov, [1,2,3,5]

[1] Skolkovo Institute of Science and Technology, Skolkovo Innovation Center 143026, 3 Nobel Street, Moscow, Russian Federation
[2] Moscow Institute of Physics and Technology, 141700, 9 Institutsky lane, Dolgoprudny, Russian Federation
[3] Dukhov Research Institute of Automatics (VNIIA), Moscow 127055, Russian Federation
[4] Institute of Physics, Jan Dlugosz University in Czestochowa, Ave. Armii Krajowej 13/15, 42-200 Czestochowa, Poland
[5] International Center for Materials Discovery, Northwestern Polytechnical University, Xi'an, 710072, China

# Crystal structures of Th-H phases

Table S1. Crystal structures of predicted Th-H phases. Volume, density and lattice parameters and coordinates are given for the lowest pressure value within the stability range.

| Phase | Pressure range, GPa | Volume per unit, Å$^3$ | $\rho$, g/cm$^3$ | Lattice parameters | Coordinates | | | |
|---|---|---|---|---|---|---|---|---|
| $C2/m$-ThH$_2$ | 0-15 | 39.78 | 9.55 | a = 4.91 Å, b = 5.95 Å, c = 3.68 Å, β = 132.14° | Th | 0.000 | 0.000 | 0.500 |
| | | | | | H | 0.000 | -0.249 | 0.000 |
| $R\bar{3}m$-ThH$_3$ | 90-100 | 24.47 | 15.35 | a = b = 2.96 Å, c = 9.67 Å, γ = 120° | Th | 0.000 | 0.000 | 0.500 |
| | | | | | H | 0.000 | 0.000 | 0.000 |
| | | | | | H | 0.000 | 0.000 | 0.289 |
| $Immm$-Th$_3$H$_{10}$ | 10-85 | 110.16 | 10.29 | a = 12.98 Å, b = 4.11 Å, c = 4.27 Å | Th | -0.331 | 0.000 | 0.000 |
| | | | | | Th | 0.000 | 0.000 | 0.000 |
| | | | | | H | 0.154 | 0.000 | -0.242 |
| | | | | | H | 0.083 | 0.500 | 0.000 |
| | | | | | H | -0.232 | 0.500 | 0.000 |
| | | | | | H | 0.000 | -0.254 | 0.500 |
| $Pnma$-ThH$_4$ | 5-10 | 44.68 | 8.77 | a = 6.65 Å, b = 4.23 Å, c = 6.35 Å | Th | 0.263 | 0.250 | 0.094 |
| | | | | | H | 0.436 | -0.018 | -0.156 |
| | | | | | H | -0.302 | 0.250 | -0.242 |
| | | | | | H | 0.391 | 0.250 | 0.443 |
| $P321$-ThH$_4$ | 15-85 | 38.34 | 10.83 | a = b = 5.96 Å, c = 3.73 Å, γ = 120° | Th | 0.000 | 0.000 | 0.500 |
| | | | | | Th | 0.333 | 0.667 | 0.210 |
| | | | | | H | 0.355 | 0.000 | 0.500 |
| | | | | | H | 0.239 | 0.000 | 0.000 |
| | | | | | H | -0.086 | -0.420 | 0.270 |
| $I4/mmm$-ThH$_4$ | 90-200 | 27.11 | 14.46 | a = b = 3.00 Å, c = 6.02 Å, | Th | 0.000 | 0.000 | 0.000 |
| | | | | | H | 0.000 | 0.500 | 0.250 |
| | | | | | H | 0.000 | 0.000 | 0.364 |
| $Cmc2_1$-ThH$_6$ | 25-90 | 41.64 | 9.49 | a = 3.99 Å, b = 6.55 Å, c = 6.38 Å | Th | 0.000 | -0.346 | -0.347 |
| | | | | | H | 0.000 | -0.086 | 0.393 |
| | | | | | H | 0.000 | 0.002 | 0.115 |
| | | | | | H | 0.000 | -0.320 | 0.029 |
| | | | | | H | 0.260 | 0.093 | 0.347 |
| | | | | | H | 0.000 | 0.201 | -0.164 |
| $P2_1/c$-ThH$_7$ | 85-115 | 33.94 | 11.69 | a = 6.09 Å, b = 3.93 Å, c = 5.78 Å, β = 78.23° | Th | -0.204 | 0.000 | 0.263 |
| | | | | | H | 0.088 | 0.224 | 0.088 |
| | | | | | H | 0.000 | -0.257 | 0.500 |
| | | | | | H | 0.303 | 0.000 | 0.105 |
| | | | | | H | 0.433 | 0.000 | 0.147 |
| | | | | | H | -0.425 | 0.000 | -0.387 |
| | | | | | H | -0.130 | 0.000 | -0.373 |
| $I\bar{4}3d$-Th$_4$H$_{15}$ | 0-10 | 188.75 | 8.29 | a = b = c = 9.04 Å | Th | 0.292 | 0.292 | 0.292 |
| | | | | | H | -0.126 | 0.222 | 0.094 |
| | | | | | H | 0.875 | 0.000 | 0.250 |
| $Fm\bar{3}m$-ThH$_{10}$ | 100-200 | 38.48 | 10.45 | a = b = c = 5.22 Å | Th | 0.000 | 0.000 | 0.000 |
| | | | | | H | 0.250 | 0.250 | 0.250 |
| | | | | | H | -0.377 | -0.377 | -0.377 |

It can be seen from Fig. S1a that $R\bar{3}m$-ThH$_3$ phase has a layered structure, where Th atoms occupy (0,0,0) positions and hydrogen atoms are in between thorium layers. Coordination number of thorium in this structure equals to 14 and distance between Th and H (d(Th-H)) equals to 2.1 Å at 90 GPa (see Fig. S1a). The shortest distance between hydrogen atoms in the layer is 1.8 Å, so they do not form a chemical bond in $R\bar{3}m$-ThH$_3$. Layered structure of $R\bar{3}m$-ThH$_3$ phase directly impacts on the electronic properties. Flat bands are seen in the Γ→A, L→M and K→H directions, which are perpendicular to the layers. The electronic density at Fermi level mainly comes from thorium atoms (red bands in Fig. S1a). The crystal structure of $Immm$-Th$_3$H$_{10}$ phase contains 12-coordinate thorium atoms in the bcc arrangement (with Th-H distance of 2.28 Å at 10 GPa) and hydrogen atoms laying along the [1,0,0], [0,1,0] and [0,0,1] planes (see Fig. S1b). ThH$_4$ phase has orthorhombic lattice with thorium atoms in (0,0,0) positions. Each thorium atom is surrounded by 12 hydrogen atoms (see Fig. S1c). The H-H distance varies from 1.57 to 1.65 Å.

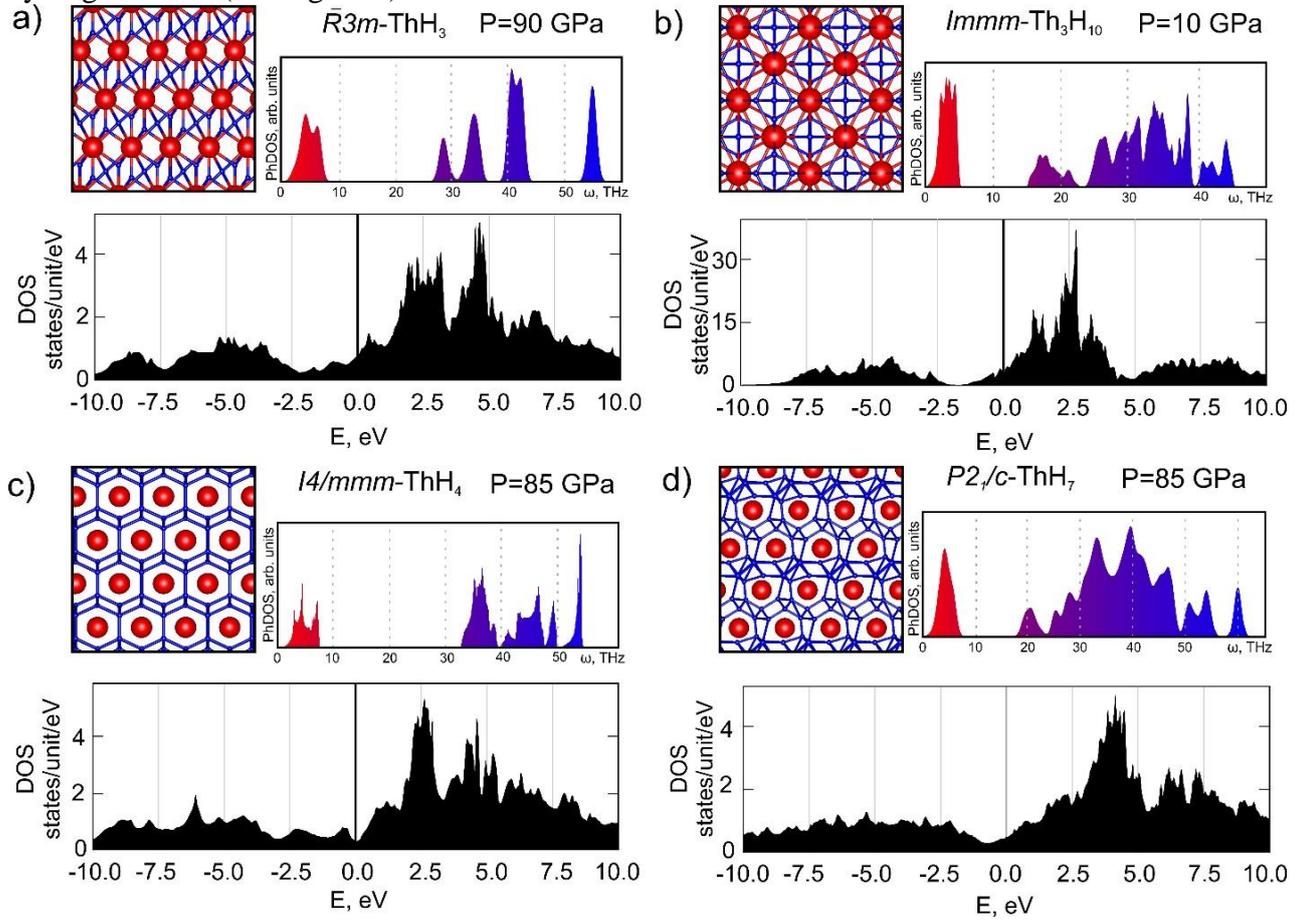

Fig. S1. Crystal structure, phonon density of states, electronic band structure and density of states for predicted a) $R\bar{3}m$-ThH$_3$, b) $Immm$-Th$_3$H$_{10}$ c) $I4/mmm$-ThH$_4$ and d) $P2_1/c$-ThH$_7$.

**Equations for calculating $T_C$**

Calculations were made using the Eliashberg equations which take the following form (on the imaginary axis) [1]:

$$\phi_n = \frac{\pi}{\beta} \sum_{m=-M}^{M} \frac{\lambda(i\omega_n - i\omega_m) - \mu^*\theta(\omega_c - |\omega_m|)}{\sqrt{\omega_m^2 Z_m^2 + \phi_m^2}} \phi_m \tag{S1}$$

$$Z_n = 1 + \frac{1}{\omega_n} \frac{\pi}{\beta} \sum_{m=-M}^{M} \frac{\lambda(i\omega_n - i\omega_m)}{\sqrt{\omega_m^2 Z_m^2 + \phi_m^2}} \omega_m Z_m \tag{S2}$$

where: $\phi_n$ – the order parameter function, $Z_n$ - the wave function renormalization factor, $\theta(x)$ - Heaviside function, $\omega_n = \pi \cdot k_B T \cdot (2n-1)$ - the n-th Matsubara frequency, $\beta = k_B T$, $\mu^*$ - Coulomb pseudopotential, $\omega_c$ - the cut-off energy ($\omega_c = 3\Omega_{max}$, $\Omega_{max}$ – maximum frequency in $\alpha^2 F(\omega)$). The electron-phonon pairing kernel:

$$\lambda(z) = 2 \int_0^{\Omega_{max}} \frac{\alpha^2 F(\Omega)}{\Omega^2 - z^2} \Omega \, d\Omega \tag{S3}$$

The Eliashberg equations have been solved for 2201 Matsubara frequencies, starting from $T_0 = 10K$ (below this temperature, the solutions are not stable). The methods discussed in the papers [2–4] were used during the calculations.

The full form of the order parameter and the wave function renormalization factor on the real axis was obtained through analytical extension of the Eliashberg equations solutions from the imaginary axis, using the formula:

$$X(\omega) = \frac{p_1 + p_2\omega + \ldots + p_r\omega^{r-1}}{q_1 + q_2\omega + \ldots + q_r\omega^{r-1} + \omega^r} \tag{S4}$$

where $X \in \{\Delta; Z\}$, $r = 50$. The values of $p_j$ and $q_j$ parameters were selected in accordance with the principles presented at work [5]. The obtained results allow to calculate dimensionless parameter $R_C$ (relative jump of the specific heat), using the formula below:

$$R_C = \frac{C^S(T_C) - C^N(T_C)}{C^N(T_C)} \tag{S5}$$

Because of strong-coupling and retardation effects, parameter $R_C$ (3.21 for $ThH_{10}$ at 100 GPa) differs signicantly from BCS theory prediction, in which the constant value is equal to 1.43 [6].

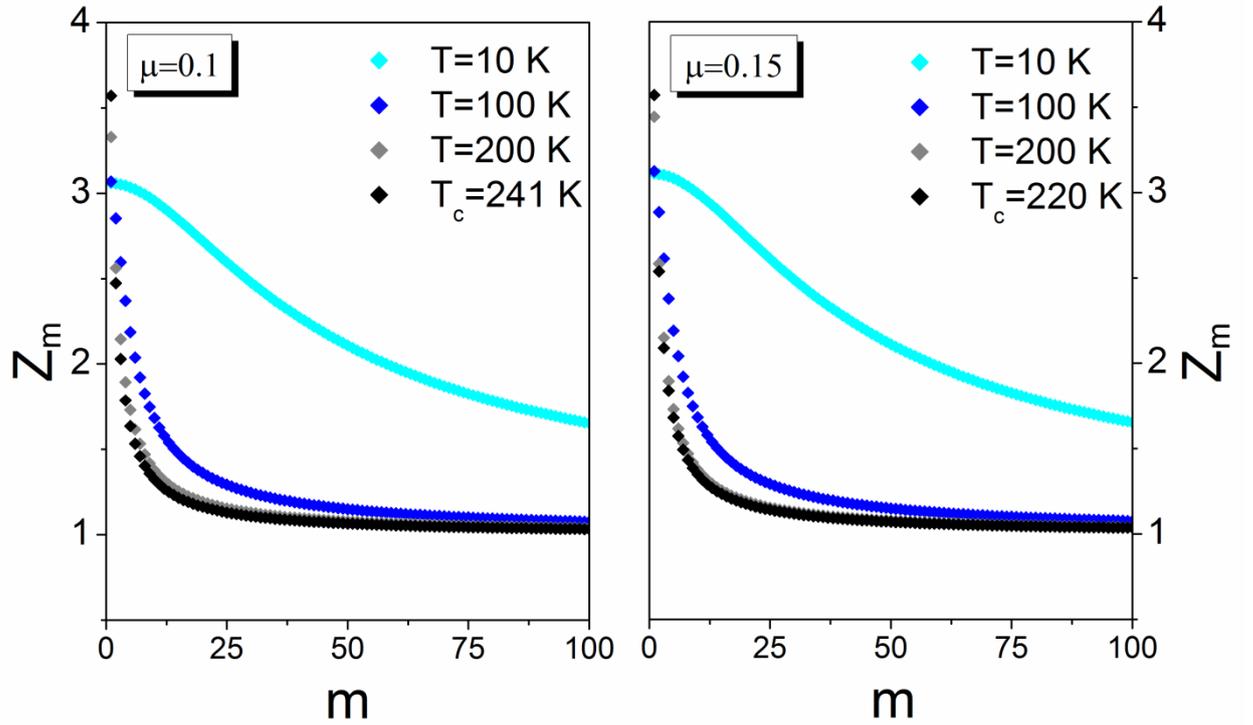

Fig. S2. The wave function renormalization factor on the imaginary axis for the selected values of temperature (the first 100 values).

Physically $Z_{m=1}$ determines approximate ratio of the renormalized electron mass ($m_e^*$) to the electron band mass ($m_e$). The wave function renormalization factor on the real axis (see Supporting Information) allows the precise calculation of the renormalized electron mass ($m_e^*$) using the formula $m_e^* = Re[Z(\omega = 0)] \cdot m_e$ [7]. The renormalized highest mass of the electron corresponds to the critical temperature, but the exact values are presented in the Table 2.

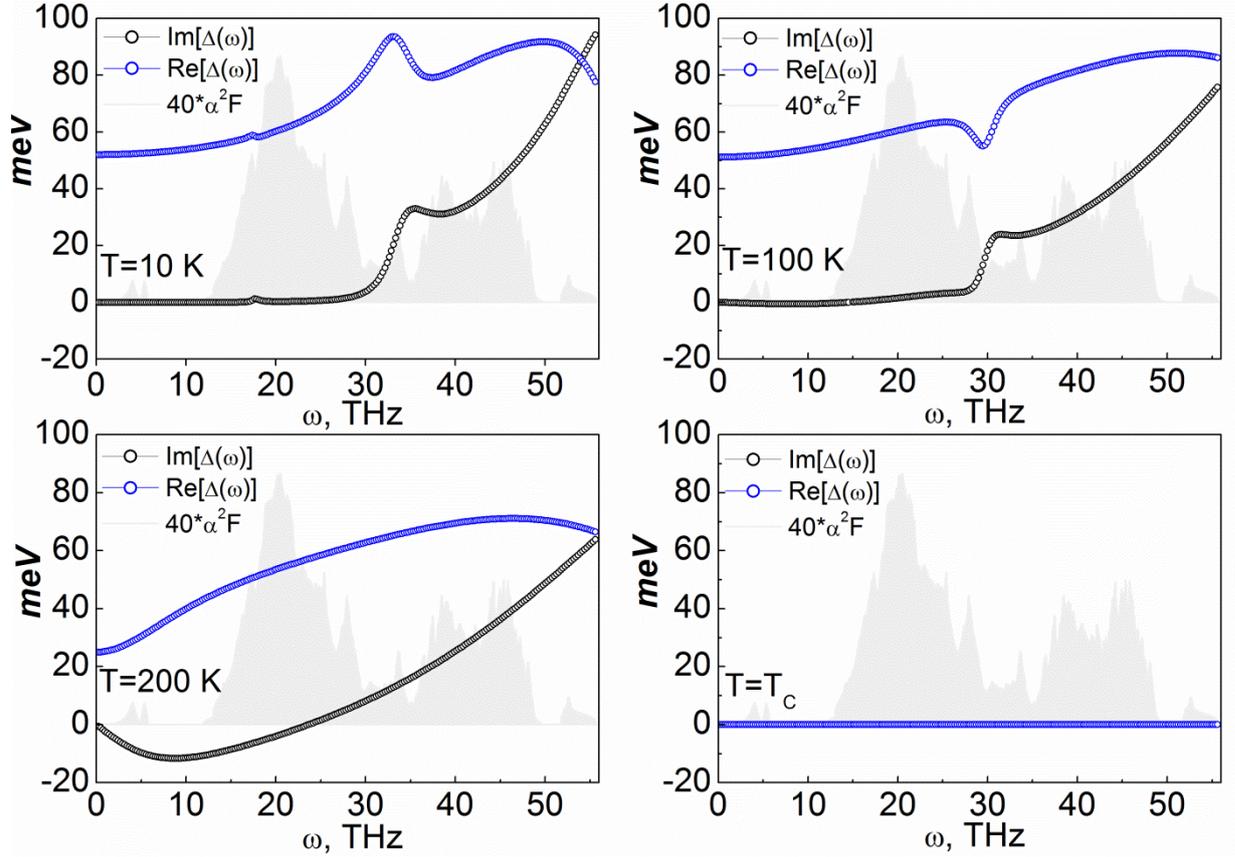

Fig. S3. The order parameter on the real axis for the selected values of temperature ($\mu* = 0.1$). In addition, the Eliashberg function (grey) was added as the background (multiplied by 40 for easier comparison).

Table S2. Parameters of superconductive state in several hydrides at optimal pressures and $\mu*=0.1$. For the cuprates and pnictides all values were obtained experimentally.

| Compound | SC gap, meV | $T_C$, K |
|---|---|---|
| $ThH_{10}$ | 52 | 241 |
| $LaH_{10}$ [8] | 68 | 286 |
| $YH_{10}$ [8] | 77 | 326 |
| $H_3S$ [9] | 42.7 | 203 |
| $BiH_6$ [10] | 18.1 | 100 |
| $PH_3$ [11] | 14.5 | 81 |
| $YH_3$ [12] | 8.4 | 45.9 |
| $H_3Se$ [13] | 28.4 | 131 |
| $MgH_6$ [14] | 106.6 | 420 |
| $YBa_2Cu_3O_{7-y}$ [15] | 34 | 92 |
| $NdBa_2Cu_3O_7$ [16] | 30 | 95 |
| $Bi_2Sr_2Ca_2Cu_3O_{10+y}$ [17] | 45 | 111 |
| $SmFeAsO_{0.9}F_{0.1}$ [18] | 15 | 44 |
| $Ba_{0.6}K_{0.4}Fe_2As_2$ [19] | 12 | 37 |

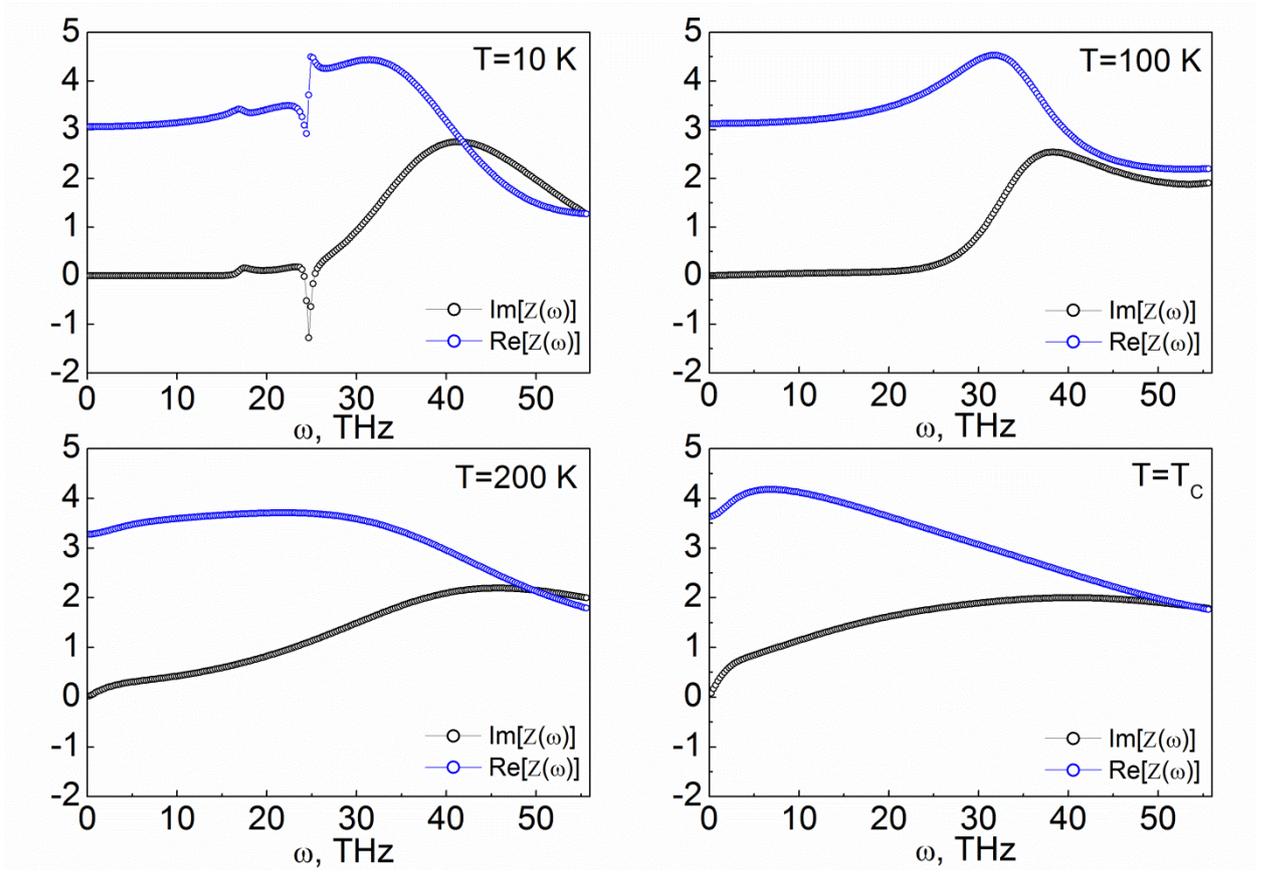

Fig. S4. The wave function renormalization factor on the real axis for the selected values of temperature ($\mu* = 0.1$). Instability of the solution at T = 10 K.

Table S3. Parameters of superconducting state of $Fm\bar{3}m$-ThH$_{10}$ at 100-300 GPa. Here $\gamma$ is Sommerfeld constant, $\mu^*$ is 0.1 (0.15).

| Parameter | $Fm\bar{3}m$-ThH$_{10}$ | | |
|---|---|---|---|
| | *100 GPa* | *200 GPa* | *300 GPa* |
| $\lambda$ | 2.50 | 1.35 | 1.11 |
| $\omega_{log}$, K | 1073.1 | 1627.1 | 1775.3 |
| $T_C$, K | 241 (220) | 228 (205) | 201 (174) |
| $\Delta(0)$, meV | 52 (47) | 45 (40) | 38 (32) |
| $H_C(0)$, T | 71 (65) | 50 (44) | 39 (33) |
| $\gamma$, mJ/mol·K$^2$ | 11 | 6.4 | 5.2 |

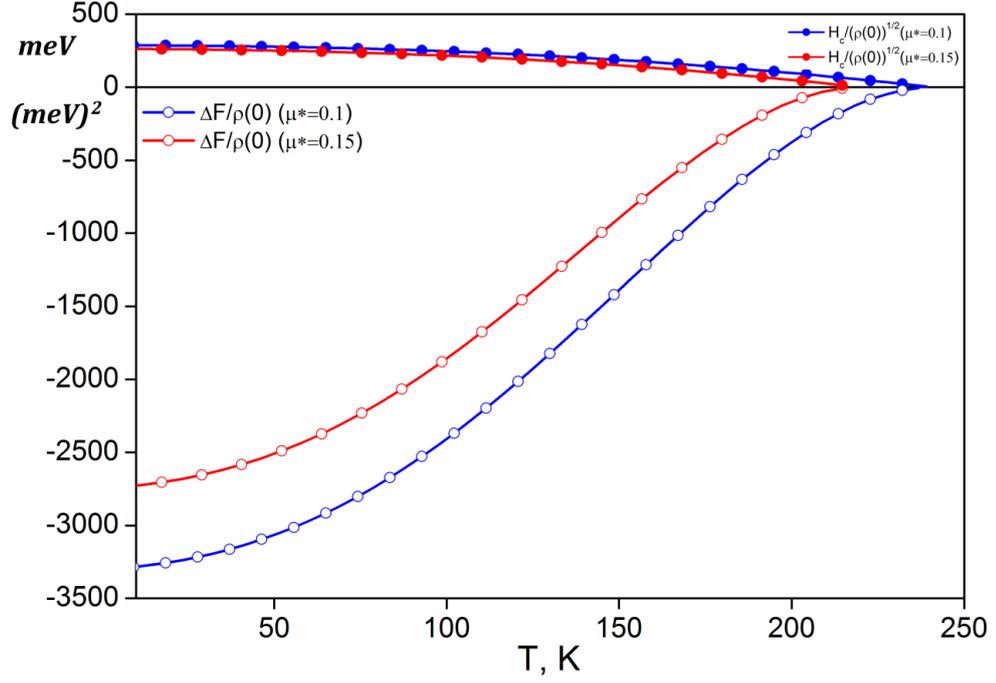

Fig. S5. Upper panel: the dependence of the thermodynamic critical field on the temperature. Lower panel: the free energy difference between the superconducting state and the normal state as a function of the temperature. The negative values of ΔF prove the thermodynamic stability of the superconducting state. For the lowest temperature, that was taken into account, the free energy difference equals 3281.35 meV$^2$ and 2724.56 meV$^2$ for $\mu^* = 0.1$ and $\mu^* = 0.15$ respectively.

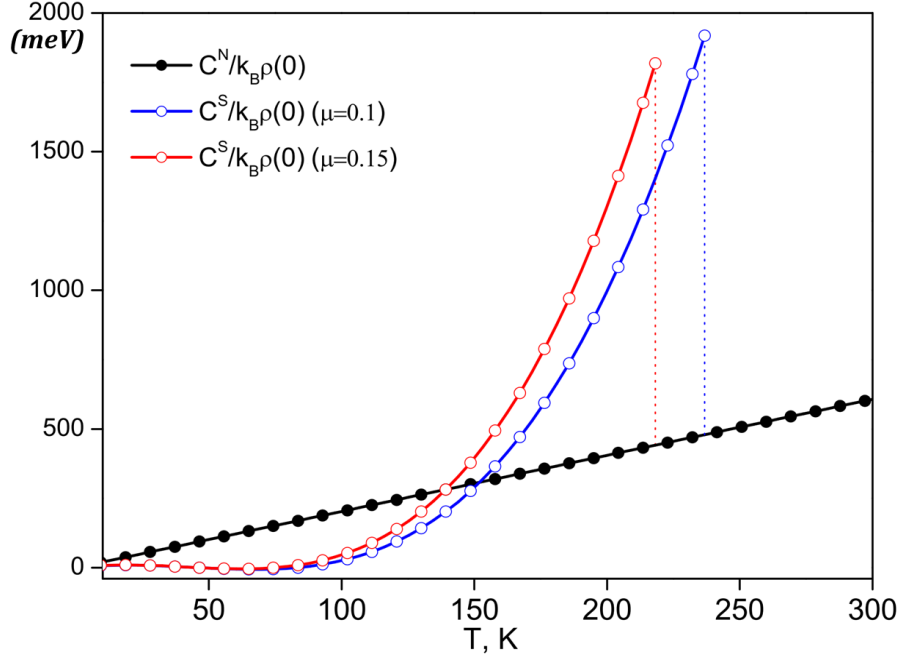

Fig. S6. The specific heat of the superconducting state ($C_S$) and the normal state ($C_N$) as a function of the temperature.

# Electronic properties of Th-H phases

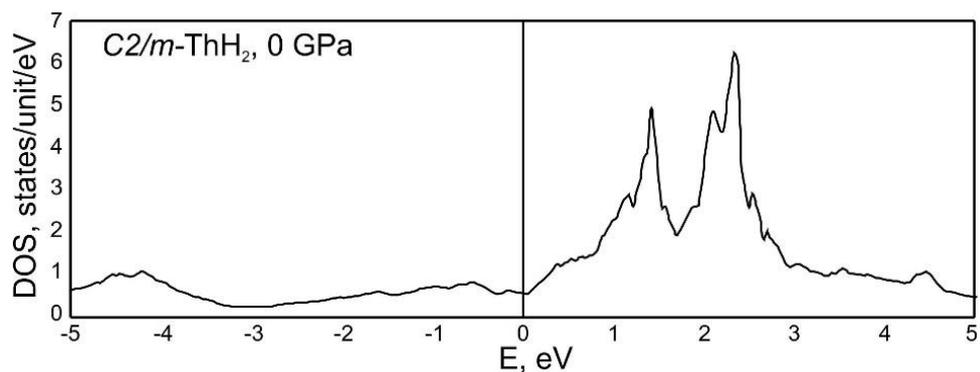

Fig. S7. Electronic density of states of $C2/m$-ThH$_2$ at 0 GPa. DOS($E_F$) = 0.35 states/unit/eV

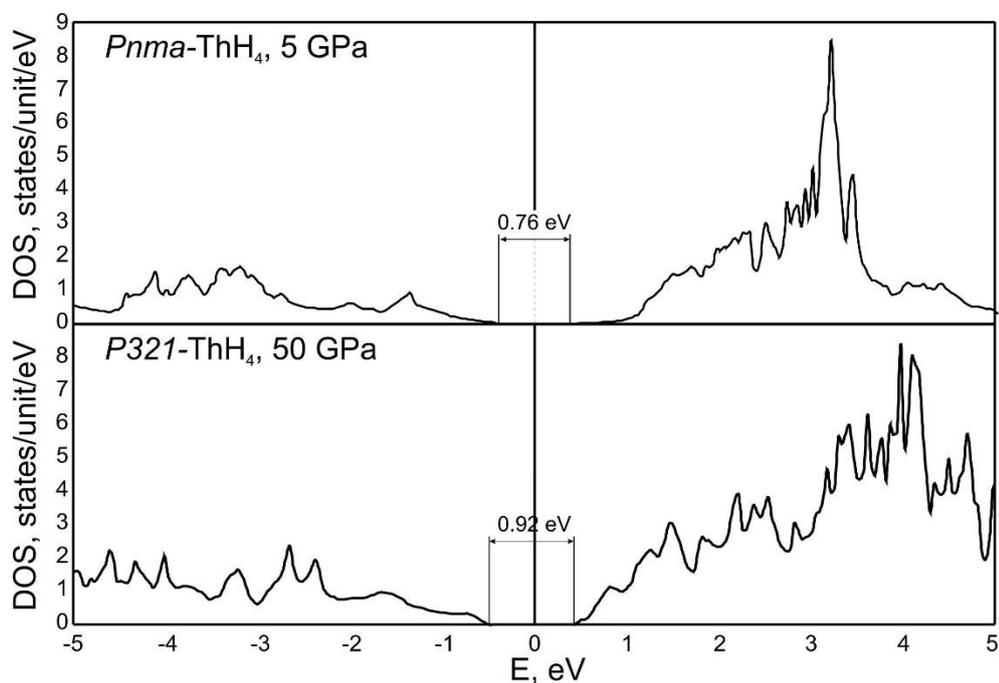

Fig. S8. Electronic densities of states of $Pnma$-ThH$_4$ and $P321$-ThH$_4$ at 5 and 50 GPa, respectively.

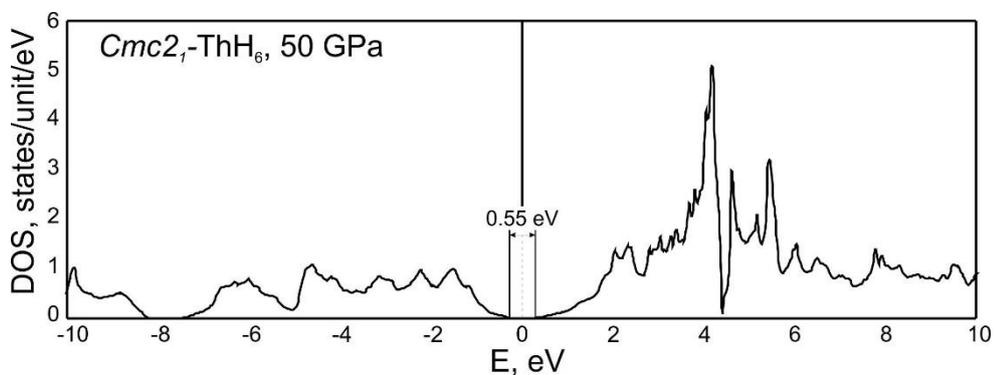

Fig. S9. Electronic density of states of $Cmc2_1$-ThH$_6$ at 50 GPa.

**Eliashberg spectral functions for ThH$_3$, Th$_3$H$_{10}$ and ThH$_7$ phases**

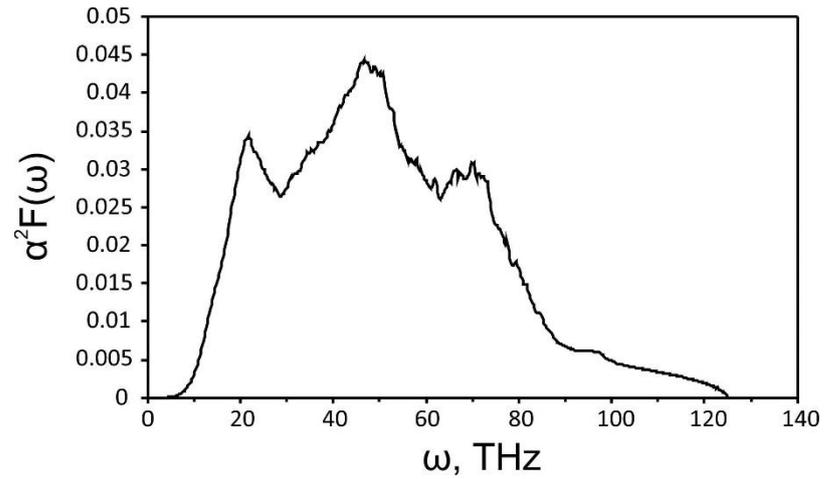

Fig. S10. Eliashberg function $\alpha^2F(\omega)$ of $R\bar{3}m$-ThH$_3$ at 100 GPa

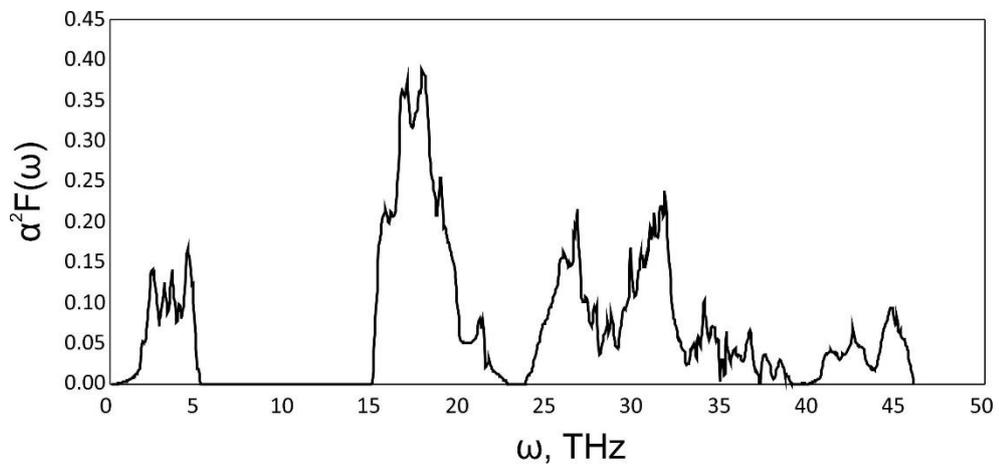

Fig. S11. Eliashberg function $\alpha^2F(\omega)$ of $Immm$-Th$_3$H$_{10}$ at 10 GPa

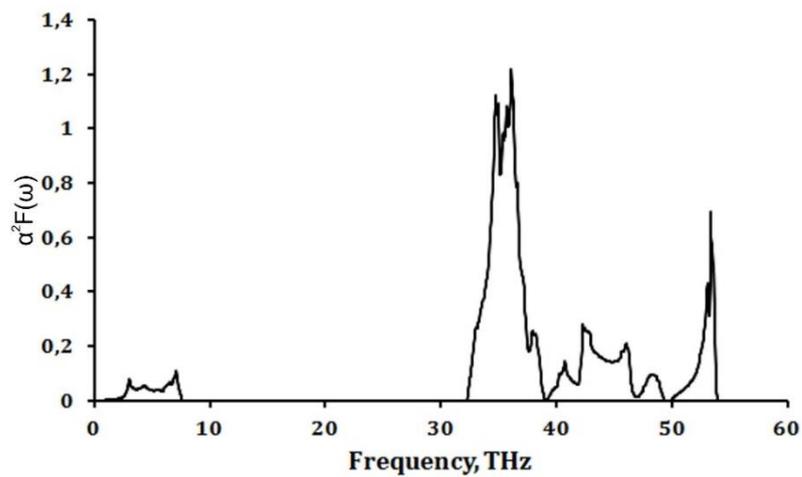

Fig. S12. Eliashberg function $\alpha^2F(\omega)$ of $I4/mmm$-ThH$_4$ at 85 GPa

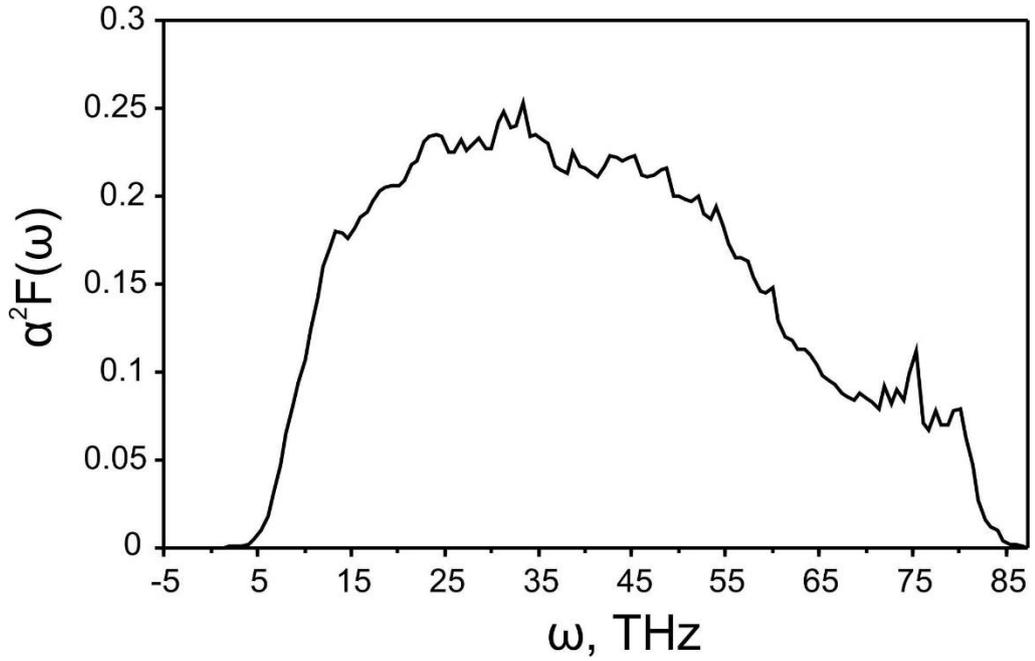

Fig. S13. Eliashberg function $\alpha^2 F(\omega)$ of $P2_1/c$-ThH$_7$ at 100 GPa

**Dependence of $T_C$ on the pressure for ThH$_{10}$**

Here we describe a detailed analysis of pressure dependence of $T_C$ of ThH$_{10}$ phase. Fig. S14 shows the electronic DOS of ThH$_{10}$ calculated at 100, 200 and 300 GPa. All the data for analysis are summarized in Table S4. It is clearly seen from both Table S4 and Fig. S14 that the density of states at Fermi level decreases linearly with pressure for ThH$_{10}$.

Table S4. Calculated data for ThH$_{10}$

| | ThH$_{10}$ | | | | | |
|---|---|---|---|---|---|---|
| P, GPa | Cell volume, Å$^3$ | $N_f$, states/unit/eV | $\omega_{log}$, K | $\lambda$ | $T_C$ (A-D)*, K | $\ln[T_C/\omega_{log}]$ |
| 100 | 248.8 | 0.61 | 1073.1 | 2.50 | 221.1 | -1.579 |
| 200 | 205.4 | 0.53 | 1627.1 | 1.35 | 182.6 | -2.187 |
| 300 | 183.2 | 0.48 | 1775.3 | 1.11 | 155.4 | -2.436 |

*Full Allen-Dynes (eq. (1))

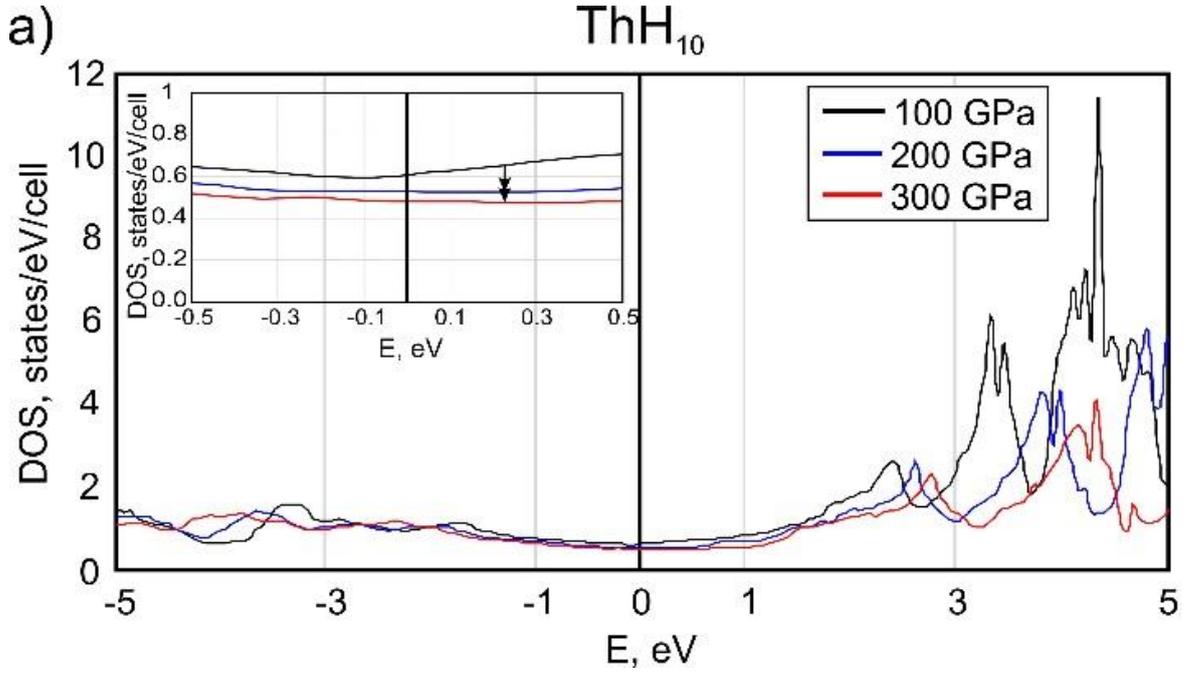

Fig. S14. Electronic densities of states of ThH$_{10}$ as a function of pressure. Fermi level is set to zero. Inset shows zoomed region from -0.5 to 0.5 eV for better view of DOS at Fermi level.

Results for ThH$_{10}$ show linear dependence of logarithmic frequency on pressure which can be well described by the following formula:

$$\gamma = \left(\frac{\partial \ln(\omega)}{\partial \ln(V)}\right)_T, \tag{S6}$$

where $\gamma$ is the Grüneisen parameter, showing only small dependence on pressure and for most materials equal to ~1-1.5.

Calculated gradient of the critical temperature dependence $dT_C/dP$ for ThH$_{10}$ phase is negative as for the majority of metals and equals to $-3.85 \cdot 10^{-5}$ K/atm at 100 GPa. This value is similar to values for low-$T_C$ superconducting metals like Hg, Ga, Bi-II [20]. This value is also similar to that of CaH$_6$ phase from Ref. [21] where the $dT_C/dP = -3.34 \cdot 10^{-5}$ K/atm. Similarity of pressure dependence of $T_C$ for ThH$_{10}$ with other metals opens a new way of using the well-known empirical equation for low-$T_C$ superconductors:

$$-\ln\left(\frac{T_C}{\omega_{log}}\right) = Cv^{-\varphi}, C > 0 \tag{S7}$$

with constant $C$ and $\varphi$ calculated earlier for non-transition metals ($\varphi = 2.5\pm0.6$) [22,23]. From calculated data (interpolation of $\ln[T_c/\omega_{log}]$ function by $a \times v^b$ law) we can directly determine the coefficients $\varphi = 1.446$, $C = 4650.1$ ($R^2 = 0.98$), see Table S4.

Using the modified McMillan equation (see eq. (2)) we derived the dependence of the EPC coefficient on the pressure as:

$$T_C = \frac{\omega_{log}}{1.2} \exp\left(\frac{-1.04(1+\lambda)}{\lambda - \mu^* - 0.62\lambda\mu^*}\right)$$

$$\ln\left(\frac{T_C}{\omega_{log}}\right) = \ln\left(\frac{1}{1.2}\right) + \left(\frac{-1.04(1+\lambda)}{\lambda - \mu^* - 0.62\lambda\mu^*}\right)\bigg|_{\mu^*=0.1} \to 0.1823 + \frac{1.04(1+\lambda)}{0.938\lambda - 0.1} \tag{S8}$$

$$= Cv^{-\varphi}$$

As a result, we obtained the following equation:

$$\lambda(P) = 0.1066 \times \frac{Cv^{-\varphi} + 10.22}{Cv^{-\varphi} - 1.291} \qquad (S9)$$

For better comparison we summarized the obtained data in the Table S5 and show it in Fig. S15.

Table S5. Comparison of numerical and analytical results for ThH$_{10}$

| P, GPa | T$_C$ (A-D), K | T$_C$ (eq. (S7)), K | λ (QE) | λ (eq. (S9)) (μ$^*$ = 0.1, φ(McM) = 1.35)$^*$ |
|---|---|---|---|---|
| 100 | 221.1 | 219.4 | 2.50 | 2.29 |
| 200 | 182.6 | 187.9 | 1.35 | 1.47 |
| 300 | 155.4 | 141.2 | 1.11 | 1.03 |

*φ(McM) – is φ based on modified McMillan equation for T$_C$.